\documentclass[twocolumn,showpacs,preprintnumbers,amsmath,amssymb]{revtex4}
\usepackage[hypertex]{hyperref}
\usepackage{times}
\usepackage{graphicx}% Include figure files
\usepackage{dcolumn}% Align table columns on decimal point
\usepackage{bm}% bold math

\newcommand{\Expect}[1]{\left\langle{#1}\right\rangle}
\newcommand{\bvec}[1]{\boldsymbol{#1}}
\newcommand{\no}{\nonumber}

\newcommand{\Integer}{\mathbb{Z}}
\newcommand{\Real}{\mathbb{R}}

\newcommand{\Complex}{\mathbb{C}}

\newcommand{\bra}[1]{\langle #1 |}
\newcommand{\ket}[1]{| #1 \rangle}

\newcommand{\U}{{\rm U}}

\begin{document}
\preprint{%\heplat{0608xxx}\\
                     UT-KOMABA/10-9\\December 2010}
\preprint{IPMU10-0231}

\title{Reflection Positivity of ${\cal N}=1$ Wess-Zumino model \\  
on the lattice with exact U(1)$_R$ symmetry}% Force line breaks with \\
\author{Yoshio Kikukawa}
	\altaffiliation{Institute of Physics, University of Tokyo, Tokyo, 153-8902, Japan}
	\email{kikukawa@hep1.c.u-tokyo.ac.jp}
\author{Kouta Usui}
	\altaffiliation{Department of physics, University of Tokyo, 113-0033, Japan\\
Institute for the Physics and Mathematics of the Universe (IPMU),   the University of Tokyo, Chiba 277-8568, Japan}
\email{kouta@hep-th.phys.s.u-tokyo.ac.jp}

%\author{Ann  Author}
%\affiliation{%

%\affiliation{
%Second institution and/or address\\%

\date{\today}% It is always \today, today,

\begin{abstract}
By using overlap Majorana fermions,  
the ${\cal N}=1$ chiral multiple can be formulated 
so that the supersymmetry is manifest and  the vacuum energy  is cancelled in the free limit, 
thanks to the bilinear nature of the free action. 
It is pointed out, however, that in this formulation %in the case of the overlap Majorana fermion, 
the reflection positivity seems to be violated in the bosonic part of the action, although it is satisfied in the fermionic part.
It is found that the positivity of the spectral density of the bosonic 
two-point correlation function is ensured only  for the spacial momenta $a | p_k | \lesssim  1.84$ $(k=1,2,3)$. 
%and the mode with a negative density appears at the energy as 
%low as $E \simeq 0.69$ for $a p_1 =\pi, ap_2=ap_3=0$.
It is then argued that 
%a possible way out of this problem
%we argue that 
%because of this fact,  
in formulating ${\cal N}=1$ Wess-Zumino model with the overlap Majorana fermion, 
%it may be better 
one may adopt  a simpler nearest-neighbor bosonic action, 
discarding the free limit manifest supersymmetry. 
The model still preserves the would-be U(1)$_R$ symmetry and satisfies the reflection positivity.
%
%Using overlap Majorana fermions,  
%the ${\cal N}=1$ chiral multiple can be formulated 
%so that the vacuum energy  is cancelled and the supersymmetry is manifest in the free limit, 
%thanks to the bilinear nature of the free action. 
%It is pointed out, however, that in this formulation %in the case of the overlap Majorana fermion, 
%the reflection positivity is violated in the bosonic part of the action, although it is satisfied in the fermionic part.
%Because of this fact,  we argue that 
%in formulating the lattice ${\cal N}=1$ Wess-Zumino model with the overlap Majorana fermion, 
%it may be better to adopt  the simpler bosonic action, 
%discarding the free limit manifest supersymmetry. 
%The model still preserves the U(1)$_R$ symmetry and satisfies the reflection positivity.

\end{abstract}

\pacs{Valid PACS appear here}% PACS, the Physics and Astronomy
\maketitle

\section{Introduction}

% supersymmetry in free limit   --> chiral multiplet 
The chiral multiplet of ${\cal N}=1$ supersymmetry\cite{Wess:1973kz} can be formulated on the lattice 
so that the supersymmetry is preserved and  the vacuum energy  is cancelled in the free limit, 
thanks to the bilinear nature of the free action.
By using overlap (Majorana) fermion \cite{Neuberger:1997fp, Neuberger:1998wv, Kikukawa:1997qh} 
for the fermionic component,  species doublers\cite{Karsten:1980wd, Nielsen:1980rz, Nielsen:1981xu} 
are successfully removed and 
U(1)$_R$ symmetry can be maintained at the same time\cite{Ginsparg:1981bj, Luscher:1998pqa, Aoyama:1998in}. 
%Here, the Ginsparg-Wilson (GW) relation\cite{Ginsparg:1981bj} leads to U(1) chiral symmetry 
%of the fermion component\cite{Luscher:1998pqa}, and 
%a bosonic counterpart of the GW relation, that emerges by imposing supersymmetry, leads to two bosonic U(1) symmetries\cite{Aoyama:1998in}. 
With this chiral multiplet, 
one may  formulate  lattice ${\cal N}=1$ Wess-Zumino model 
with exact U(1)$_R$ symmetry\cite{Fujikawa:2001ka,Fujikawa:2001ns, Fujikawa:2002ic, Bonini:2004pm, Kikukawa:2004dd,Bonini:2005qx}.
A numerical study of this lattice ${\cal N}=1$ Wess-Zumino model % with exact U(1)$_R$ symmetry
has recently been reported in \cite{Chen:2010uc}. 

The purpose of this short article is, however,  to show that in this formulation of the chiral multiplet,  
%in the case of the overlap Majorana fermion, 
the reflection positivity\cite{Osterwalder:1973dx, Osterwalder:1974tc, Osterwalder:1977pc, Luscher:1976ms,
Menotti:1987cq} seems to be violated in the bosonic part of the action, although it is satisfied in the fermionic part, 
as shown recently in \cite{Kikukawa:2010gq}.
We will also examine the spectral density of the bosonic two-point correlation function (cf. \cite{Luscher:2000hn}). 
%and 
%identify the energy-momentum region where the spectral density becomes negative (cf. \cite{Luscher:2000hn}). 
It is found that the positivity of the spectral density 
%of the bosonic two-point correlation function 
is ensured only  for the momenta $a | p_k | \lesssim  1.84$ $(k=1,2,3)$, and
the mode with a negative density appears at the energy as low as $a E \simeq 0.69$ for the momenta
$a \bvec{p} =(\pi, 0, 0),(0,\pi,0),(0,0,\pi)$.
% for 
%\begin{equation}
%a p \ge  1.7 ,  \quad a E \simeq  1.7 . 
%\end{equation}

We will  then argue that 
in formulating the lattice ${\cal N}=1$ Wess-Zumino model with the overlap (Majorana) fermion, 
one may adopt  the simpler nearest-neighbor bosonic action, discarding the free limit manifest supersymmetry. 
The model so constructed still preserves the U(1)$_R$ symmetry and satisfies the reflection positivity.

This paper is organized as follows.  
In section~\ref{sec:overlap-chiral-multiplet}, we review briefly the ${\cal N}=1$ chiral multiple on the lattice 
formulated with overlap Majorana fermion.
In section~\ref{sec:RF-violation-bosonic-part}, we show that the standard
way to prove the reflection positivity does not work in the 
bosonic part of the action.   The spectral density of the bosonic two-point correlation function is also examined.
In section~\ref{sec:RF-lattice-WZ-model},  we show that it is possible to formulate lattice Wess-Zumino model 
which possesses both the reflection positivity and the exact U(1)$_R$ symmetry, 
by adopting the simpler nearest-neighbor bosonic action. 
Section~\ref{sec:discussion} is devoted to discussion.

\section{${\cal N}=1$ chiral multiple with overlap Majorana fermion}
\label{sec:overlap-chiral-multiplet}

%By using the overlap Majorana fermion, it is possible to formulate ${\cal N}=1$ chiral multiple which 
%possesses  ${\cal N}=1$ supersymmetry and  U(1)$_R$ symmetry manifestly in the free limit. 
The action of the free ${\cal N}=1$ chiral multiplet  is given by
\begin{eqnarray}\label{chiral multiplet}
   S_0&=&a^4\sum_x\Bigl\{ \, 
   {1\over2}\chi^TCD_1\chi +\phi^*D_1^2\phi+F^*F   \nonumber\\
   &&\qquad
    +{1\over2}\chi^TCD_2 \chi +FD_2\phi+F^*D_2\phi^*   \, \Bigr\} . 
%\nonumber\\
%   &&\qquad\qquad\qquad\quad
%   -{1\over a}X^TCX
%   -{2\over a}(\mathcal{F}\Phi+\mathcal{F}^*\Phi^*)
%   \, \Bigr\},
\end{eqnarray}
%Here,  we adopt the overlap-Dirac operator~\cite{Neuberger:1998fp} defined
%by\footnote{$\partial_\mu f(x)=\{f(x+a\hat\mu)-f(x)\}/a$
%and~$\partial_\mu^*f(x)=\{f(x)-f(x-a\hat\mu)\}/a$ are the forward and backward
%difference operators, respectively.}
%\begin{equation}
%   D={1\over2}\{1-A(A^\dagger A)^{-1/2}\},\qquad
%   A=1-aD_{\rm w},\qquad
%   D_{\rm w}={1\over2}\{\gamma_\mu(\partial_\mu^*+\partial_\mu)
%   -a\partial_\mu^*\partial_\mu\},
%\end{equation}
%which obeys the Ginsparg-Wilson
%relation~$\gamma_5D+D\gamma_5=aD\gamma_5D$~\cite{Ginsparg:1982bj}. 
In this expression,  we have
used a decomposition of the overlap Dirac operator~\cite{Neuberger:1997fp}\cite{Neuberger:1998wv}, $D=D_1+D_2$, where
\begin{align}
   D_1&={1\over2}\gamma_\mu(\partial_\mu^*+\partial_\mu)(A^\dagger A)^{-1/2},\\
   D_2&={1\over a}\Bigl\{1-
   (1+{1\over2}a^2\partial_\mu^*\partial_\mu)(A^\dagger A)^{-1/2}\Bigr\},
\label{twoxten}
\end{align}
and
\begin{align}
A=1-aD_{\text{w}},
\quad 
D_{\text{w}}=\frac{1}{2}\Bigl\{\gamma_\mu(\partial_\mu^*+\partial_\mu)-a\partial_\mu^*\partial_\mu\Bigr\}.
\end{align}
Note that $D_1$ and~$D_2$ have different spin 
structures with respect to spinor space. In particular, we have $\{\gamma_5,D_1\}=0$ and~$[\gamma_5,D_2]=0$.
In terms of this decomposition, the Ginsparg-Wilson relation~$\gamma_5D+D\gamma_5=aD\gamma_5D$~\cite{Ginsparg:1981bj}
 is expressed as
\begin{equation}
   2D_2=a(-D_1^2+D_2^2),
\label{twoxeleven}
\end{equation}
and as a consequence, we have relations
\begin{align}
   \gamma_5(1-{1\over2}aD)\gamma_5(1-{1\over2}aD)&=1-{1\over2}aD_2,\\
   \gamma_5(1-{1\over2}aD)\gamma_5D&=D_1 . 
\label{twoxtwelve}
\end{align}
It is also understood that the $4\times 4$
identity matrix in operators $D_1^2$ and~$D_2$ is omitted when these operators
are acting on bosonic fields. 

It is straightforward to see that the
above free action~$S_0$ is invariant under ``lattice ${\cal N}=1$ supersymmetry": 
\begin{eqnarray}
   &&\delta_\epsilon\chi=-\sqrt{2}P_+(D_1\phi+F)\epsilon
   -\sqrt{2}P_-(D_1\phi^*+F^*)\epsilon,
\nonumber\\
   &&\delta_\epsilon\phi
   =\sqrt{2}\epsilon^TCP_+\chi,\quad
   \delta_\epsilon\phi^*=\sqrt{2}\epsilon^TCP_-\chi,
\nonumber\\
   &&\delta_\epsilon F=\sqrt{2}\epsilon^TCD_1P_+\chi,\quad
   \delta_\epsilon F^*=\sqrt{2}\epsilon^TCD_1P_-\chi,
\label{twoxeight}
\end{eqnarray}
where $\epsilon$ is a 4~component Grassmann parameter. 
We also  note that the free action~$S_0$ possesses three types of $\U(1)$
symmetry~\cite{Aoyama:1998in}. The first is a rather trivial one acting only
on bosonic fields and is defined by the transformation:
\begin{eqnarray}
   &&\delta_\alpha\chi=0, %\qquad\delta_\alpha X=0,
\nonumber\\
   &&\delta_\alpha\phi=i\alpha\phi, %\qquad \delta_\alpha\Phi=i\alpha\Phi,
\nonumber\\
   &&\delta_\alpha F=-i\alpha F, %\qquad  \delta_\alpha\mathcal{F}=-i\alpha\mathcal{F},
\label{twoxfifteen}
\end{eqnarray}
where $\alpha$ is an infinitesimal real parameter. 
%This remains the symmetry of~$S_0$ even for~$m\neq0$. 
The second one is nothing but the chiral
symmetry introduced by L\"uscher, 
\begin{eqnarray}
   &&\delta_\alpha\chi=i\alpha\gamma_5(1-{1\over2}aD)\chi , 
   %   +i\alpha\gamma_5X,
   %\qquad \delta_\alpha X=i\alpha\gamma_5{1\over2}aD\chi,
%\nonumber\\
%   &&\delta_\alpha\phi=-2i\alpha\phi,\qquad
%   \delta_\alpha\phi^*=2i\alpha\phi^*.
\label{twoxseven}
\end{eqnarray}
Thirdly, somewhat surprisingly, the {\it bosonic\/} sector of $S_0$
possesses a $\U(1)$ symmetry analogous to eq.~(\ref{twoxseven}):
\begin{eqnarray}
%   &&\delta_\alpha\chi=0,\qquad \delta_\alpha X=0,
%\nonumber\\
   &&\delta_\alpha\phi
   =+i\alpha\{(1-{1\over2}aD_2)\phi-{1\over2}aF^*\} , %+i\alpha\Phi,
%   \qquad
%   \delta_\alpha\Phi=+i\alpha\{{1\over2}aD_2\phi+{1\over2}aF^*\},
\nonumber\\
   &&\delta_\alpha F=+i\alpha\{(1-{1\over2}aD_2)F-{1\over2}aD_1^2\phi^*\} %+i\alpha\mathcal{F},
%   \qquad
%   \delta_\alpha\mathcal{F}=+i\alpha\{{1\over2}aD_2F+{1\over2}aD_1^2\phi^*\},
%\nonumber\\
\label{twoxsixteen}
\end{eqnarray}
due to the Ginsparg-Wilson relation. The lattice action~$S_0$ is not invariant
under a uniform rotation of the complex phase of bosonic fields, $\phi$, $F$,
%$\Phi$ and~$\mathcal{F}$, 
due to the presence of terms~$FD_2\phi$
and~$F^*D_2\phi^*$. The above provides a lattice counterpart of this uniform
phase rotation of bosonic fields under which the free action~$S_0$
is invariant. Using a linear combination of the above three $\U(1)$ symmetries,
it is possible to define the $\U(1)_R$ symmetry~\cite{Aoyama:1998in} in the
interacting system. %, as we will see below. 
\begin{align}
   &\delta_\alpha\chi=+i\alpha\gamma_5(1-{1\over2}aD)\chi , 
%   +i\alpha\gamma_5X,\qquad
%   \delta_\alpha X=+i\alpha\gamma_5{1\over2}aD\chi,
\nonumber\\
   &\delta_\alpha\phi
   =-3i\alpha\phi+i\alpha\{(1-{1\over2}aD_2)\phi-{1\over2}aF^*\} , 
%   +i\alpha\Phi,
\nonumber\\
%   &&\delta_\alpha\Phi=-3i\alpha\Phi
%   +i\alpha\{{1\over2}aD_2\phi+{1\over2}aF^*\},
%\nonumber\\
   &\delta_\alpha F=+3i\alpha F
   +i\alpha\{(1-{1\over2}aD_2)F-{1\over2}aD_1^2\phi^*\} . 
%   +i\alpha\mathcal{F},
%\nonumber\\
%   &&\delta_\alpha\mathcal{F}=+3i\alpha\mathcal{F}
%   +i\alpha\{{1\over2}aD_2F+{1\over2}aD_1^2\phi^*\}.
\label{twoxtwentythree}
\end{align}

\section{Violation of the reflection positivity in the bosonic part}
\label{sec:RF-violation-bosonic-part}
%It is desireble 
%that a lattice field theory is formulated in the way that it satisfies the fundamental requirements
%such as locality and unitarity (reflection positivity) as well as 
%important symmetries of the target continuum theory one of 
%which is the supersymmetry. 
%These properties help garantee that the lattice field theory is a
%suitable effective theory belonging to the same universality class as the target continuum field theory.
%and this is the very reason why we can
%trust the results obtained through the lattice field theory. 
%Therefore, in the formulation of the supersymmetric 
%field theory on the lattice, it is important to check whether these fundamental
%requirements are satisfied or not.  % completed only by keeping the supersymmetry on the lattice. 
%We should prove, in addition to the supersymmetric properties, that the locality and the
%reflection positivity are fulfilled, and this is not at all an obvious problem. 
%In this section, we discuss the reflection
%positivity of the lattice Wess Zumino model with the overlap Dirac operator, as formulated 
%in the previous section. %in which the $U(1)_R$
%symmetry and the free limit supersymmetry are kept on the lattice. 
%We will see in the following that the `overlap boson', the bosonic theory defined by the overlap Dirac operator,
%does \textit{not} satisfy the reflection positivity condition, which implies the violation of the
%reflection positivity of the Wess Zumino model. 

\subsection{Reflection positivity condition}
%The reflection positivity condition, which is a condition for the Euclidean field theory, 
%is almost equivalent to the unitarity of the Minkowski quantum filed theory. In fact,
 In this subsection, we will formulate
the reflection positivity condition. 
It has been rigorously shown that the lattice theory satisfying the reflection positivity condition   
corresponds to the quantum theory with unitary time evolution \cite{Osterwalder:1973dx, Osterwalder:1974tc, Osterwalder:1977pc}.
Here we consider the generic case in which there are both a bosonic field $\phi$ and 
a fermionic field $\psi$. Let us assume that $S(\phi,\psi,\bar\psi)$ is the action of a lattice model\footnote{
In the following, 
we write the bosonic field argument of a function like $S(\phi)$ instead of $S(\phi,\phi^*)$
for the notational simplicity. This notation never means that $S$ is an analytic function of $\phi$.
}
and its partition function $Z$ is given by the path integration
\begin{align}
Z=\int[{\cal D \phi}{\cal D \phi^*}][{\cal D\psi}{\cal D\bar{\psi}}]
\,{\rm e}^{-S(\phi,\psi,\bar\psi)}.
\end{align}
We set the lattice spacing $a$ to be unity, and assume the finite volume hypercubic lattice
$\Lambda=\{-L+1,-L+2,\dots,L-1,L\}^d\subset\Integer^d$.
We impose the \textit{anti}-periodic boundary condition in the time direction for the fermionic field $\psi$, while
the periodic boundary condition for the  bosonic field $\phi$. In the spacial directions, periodic boundary conditions
are imposed for both fields.

To formulate the reflection positivity condition, we first introduce the time reflection operator $\theta$ as 
follows. 
For each site $x=(t,\bvec x)\in\Lambda$, we denote $\theta x=(-t+1,\bvec x)$. This is the
time reflection with respect to the $t=1/2$ plane. We define the operation of $\theta$
for bosonic fields as
\begin{align}
(\theta\phi)(x)=\phi(\theta x)
\end{align} 
and for functions of bosonic fields ${\cal F}(\phi)$ as
\begin{align}\label{theta for boson}
(\theta {\cal F})(\phi)={\cal F}^*(\theta\phi),
\end{align}
where * means complex conjugation. For fermionic fields, 
the $\theta$ reflection is defined as
\begin{align}\label{theta for fermion}
(\theta\bar\psi)(x)&=\gamma_0\psi(\theta x) ,\\
(\theta\psi)(x)&=\bar\psi(\theta x)\gamma_0.
\end{align}

We extend this $\theta$ operation to the whole field algebra $\mathcal{A}$.
We define the field algebra $\mathcal{A}$, the algebra of observables, as 
the Grassmann algebra generated by the fermionic fields
with the coefficients of the continuous functions of bosonic fields which are integrable with respect to 
the bosonic Gaussian functional measure. For $\mathcal{F},\mathcal{G}\in\mathcal{A}$, 
the $\theta$ operation is defined by the relations
\begin{align}
\theta({\cal FG})&=\theta({\cal G})\theta({\cal F}) ,\label{product}\\
\theta(\alpha {\cal F} +\beta {\cal G})&=\alpha^*\theta({\cal F})+\beta^*\theta({\cal G}).\label{linear combi}
\end{align}
%for $\mathcal{F},\mathcal{G}\in{\cal A}$.
For instance, if ${\cal F}$ has the form of
\begin{align}
{\cal F(\phi,\psi,\bar\psi)}=&f(\phi)\bar\psi_{a_1}(x_1)\dots\bar\psi_{a_n}(x_n)\times\no\\
&\quad\times\psi_{b_1}(y_1)\dots\psi_{b_m}(y_m),
\end{align}
its $\theta$ reflection should be
\begin{align}
\theta({\cal F})(\phi,\psi,\bar\psi)&=f^*(\theta\phi)(\bar\psi\gamma_0)_{b_m}(\theta y_m)\dots(\bar\psi\gamma_0)_{b_1}(\theta y_1)\times\no\\
&\quad\times(\gamma_0\psi)_{a_n}(\theta x_n)\dots(\gamma_0\psi)_{a_1}(\theta x_1).
\end{align}
Let $\Lambda_{+}$ (resp. $\Lambda_-$) be the set of lattice sites with positive (resp. non-positive) time
components, and ${\cal A_\pm}$ be the subalgebras of $\mathcal{A}$, which depends only upon
fields on $\Lambda_\pm$. In this notation, $\theta$ is a map from $\Lambda_\pm$ into
$\Lambda_\mp$ and from ${\cal A_\pm}$ into ${\cal A_\mp}$.

%The reflection positivity condition is imposed on the expectation functional 
Reflection positivity condition is defined through this $\theta$ map.
For a lattice theory with the expectation functional $\Expect \cdot$
defined for $ \mathcal{F}\in{\cal A}$ as
\begin{align}
\Expect{\mathcal{F}}=\frac{1}{Z}\int[{\cal D \phi}{\cal D \phi^*}][{\cal D\psi}{\cal D\bar{\psi}}]
\,{\rm e}^{-S(\phi,\psi,\bar\psi)}\mathcal{F}(\phi,\psi,\bar\psi), 
\end{align}
 we say the theory is reflection positive
with respect to $\theta$ if any function ${\cal F}_+\in{\cal A_+}$ 
%i.e. any function $\mathcal{F}_+$ depending only on the positive time fields,
fulfills the inequality
\begin{align}\label{reflection positivity}
\Expect {\theta({\cal F}_+){\cal F}_+}\ge 0.
\end{align}

\subsection{Reflection positivity of the free overlap boson}
%In the lattice Wess Zumino model,
%the relfection positivity may be violated because of the use of the overlap Dirac operator, which is
%far more complicated than the Wilson Dirac operator. Although the reflection positivity of the
%overlap gauge theory is not proved yet, the reflection 
%positivity of the overlap Dirac and Majorana fermion is rigorously established in the case of non-gauge theory \cite{Kikukawa:2010gq}. 
%Hence, we don't have to worry the fermionic part and can concentrate on the study of the bosonic part.
 %and therefore no one knows in what sense this lattice model is a \textit{quantum} field theory
%and how one can believe that the results obtained through this model are correct.
 
In this subsection, we investigate the reflection positivity of the bosonic sector of the free chiral multiplet 
\eqref{chiral multiplet}. 
It will be shown in the following that the bosonic sector does not seem to satisfy the reflection positivity
condition.
After integrating out the auxiliary field $F$, we have
 the overlap boson system 
%Here, we consider the reflection positivity of the system of the overlap boson 
which is defined through
the lattice action on $\Lambda$
\begin{align}\label{bosonic action}
S_b(\phi)&=\sum_{x\in\Lambda}\phi^*(x)\Box_{\Lambda}\phi(x), 
\end{align}
where we have defined
\begin{align}
\Box_\Lambda(x,y)&=\sum_{n\in\Integer^4}\Box(x+2nL,y),\label{box on lambda}
%\\
%D^\dagger D&=\Box\cdot\hat{1},
\end{align}
and 
\begin{align}
D^\dagger D &= \Bigl\{1-
(1+{1\over2}\partial_\mu^*\partial_\mu)(A^\dagger A)^{-1/2}\Bigr\} \\
&=\Box\cdot\hat{1}
\end{align}
with $\hat{1}$ being the unit spinor matrix.
%and 
%\begin{align}
%\Box=\Bigl\{1-
%(1+{1\over2}\partial_\mu^*\partial_\mu)(A^\dagger A)^{-1/2}\Bigr\}.
%\end{align}
%the overlap operator $D$ is defined by using the ordinary Wilson Dirac operator
%\begin{align}
%D=\frac{1}{2}\Big(1+\frac{D_\text{w}-1}{\sqrt{(D_\text{w}-1)^\dagger(D_\text{w}-1)}}\Big).
%\end{align}
The operator $\Box$ given above is the bosonic overlap operator on $\Integer^d$
and $\Box_{\Lambda}$ is that on $\Lambda$ with periodic boundary conditions.
% The definition \eqref{box on lambda} means that we have chosen the
%periodic boundary condition.
The field algebra ${\cal A}$ of this overlap boson system is defined as the set of 
all continuous functions of bosonic field configulations $\phi=\{\phi(x)\}_{x\in\Lambda}$,
which are integrable with respect to the bosonic Gaussian measure
\begin{align}
[{\cal D\phi}][{\cal D\phi^*}]\,{\rm e}^{-S_b(\phi)}.
\end{align}
The expectation of this theory is defined by the bosonic path integration 
\begin{align}
\Expect{\mathcal{F}}=\frac{1}{Z}\int[{\cal D\phi}][{\cal D\phi^*}]\,{\rm e}^{-S_b(\phi)}\mathcal{F}(\phi), \qquad \mathcal{F}\in{\cal A}.
\end{align}
%where ${\cal A}$ here is the set of 
%the ``field algebra''.
 
The standard way of investigating the reflection positivity of lattice field theory 
is to prove that the action can be written in the form of
\begin{align}
-S_b(\phi)=B(\phi)+\theta(B)(\phi) +\sum_s\theta(C_s)(\phi)C_s(\phi)&,\label{decomposition of the action}
\end{align}
with $B,C_s\in{\cal A_+}$,
where in the third term $C_s$ are elements of ${\cal A_+}$ parametrized by some discrete parameter $s$ \cite{Osterwalder:1977pc}. 
To see that the equation \eqref{decomposition of the action} indeed implies the reflection positivity
\eqref{reflection positivity}, we first note that % \cite{Osterwalder:1977pc}.
for an arbitrary ${\cal F}_+\in{\cal A_+}$,
\begin{align}
&\Expect{\theta({\cal F}_+){\cal F}_+}_0\no\\
&=\int [{\cal D\phi}][{\cal D\phi^*}]\theta({\cal F}_+){\cal F}_+\no\\
&=\int\prod_{x\in\Lambda_+}d\phi(x)d\phi^*(x)\,{\cal F}_+(\phi)
\int\prod_{x\in\Lambda_-}d\phi(x)d\phi^*(x)\,\theta({\cal F}_+)(\phi) \no\\
&=\left| \int\prod_{x\in\Lambda_+}d\phi(x)d\phi^*(x)\,{\cal F}_+(\phi) \right|^2 \ge 0.\label{zero positive}
\end{align}
If the action is given in the form of \eqref{decomposition of the action}, we obtain for all ${\cal F}\in\mathcal{A}_+$,
\begin{align}\label{positivity of zero expectation}
&\Expect{{\rm e}^{-S_b}\theta({\cal F}_+){\cal F}_+}_0\no\\
&=\Expect{{\rm e}^{B+\theta(B) +\sum_s\theta(C_s)C_s}\theta({\cal F}_+){\cal F}_+}_0 \no\\
&=\Expect{\theta({\rm e}^{B})\,{\rm e}^{B}\,\sum_{n=0}^\infty\frac{1}{n!}\Big(\sum_s \theta(C_s)C_s\Big)^n\,\theta({\cal F}_+){\cal F}_+}_0 \no\\
&=\sum_{n=0}^{\infty}\frac{1}{n!}\sum_{s_1\dots s_n}\Expect{
\theta({\rm e}^{B})\,{\rm e}^{B} \,\theta(C_{s_1})C_{s_1}\dots\theta(C_{s_n})C_{s_n}\,\theta({\cal F}_+){\cal F}_+}_0\no\\
&=\sum_{n=0}^{\infty}\frac{1}{n!}\sum_{s_1\dots s_n}\Expect{
\theta\Big({\rm e}^{B}C_{s_1}\dots C_{s_n}{\cal F}_+\Big)\,{\rm e}^{B} C_{s_1}\dots C_{s_n}{\cal F}_+\Big)}_0.
\end{align}
This last expression is clearly positive from \eqref{zero positive}. This immediately implies the reflection positivity because
\begin{align}\label{RP}
\Expect{\theta({\cal F}_+){\cal F}_+}
=\frac{\Expect{{\rm e}^{-S_b}\theta({\cal F}_+){\cal F}_+}_0}{\Expect{{\rm e}^{-S_b}}_0}\ge 0, \qquad {\cal F}_+\in{\cal A}_+.
\end{align}
We note that the third term in \eqref{decomposition of the action} may be given by an integration
over a continuous parameter $s$ as
 \begin{align}
 \int ds\, \theta(C_s)(\phi)C_s(\phi),\qquad C_s\in\mathcal{A}_+.
 \end{align}
 This type of the action appears in the case of overlap fermions. See Ref.\cite{Kikukawa:2010gq} for detail.

Therefore, to prove the reflection positivity of the `overlap boson' system reduces to find
the decomposition of the action \eqref{bosonic action} into \eqref{decomposition of the action}. % or \eqref{decomposition of the action conti}. 
We first note that $S_b$ can be written as
\begin{align}
S_b=&\sum_{x,y\in\Lambda_+}\phi^*(x)\Box(x,y)\phi(y)
+\sum_{x,y\in\Lambda_-}\phi^*(x)\Box(x,y)\phi(y)\no\\
&\qquad+2\sum_{x\in\Lambda_+,y\in\Lambda_-}\phi^*(x)\Box(x,y)\phi(y),
\end{align}
where $\Box(x,y)$ is the kernel of the operator $\Box$ on $\Lambda$. 
To establish the decomposition \eqref{decomposition of the action}, we should find that  (i) the second term 
is the $\theta$ reflection of the first term, and that (ii) the last term is written in the form of %$-\sum_{j}\theta(C_j)C_j$
%for some $C_j\in{\cal A_+}$, or in the form of 
\begin{align}
-\int\theta(C_s)C_s\,ds
\end{align}
 for some $C_s\in\mathcal{A}_+$ parametrized by some parameter $s$. Note that this second condition is equivalent to say that
 \begin{align}\label{reflection property}
-\sum_{x\in\Lambda_+,y\in\Lambda_-}\phi^*(x)\Box(x,y)\phi(y)=\int f(s)\,\theta(C_s)(\phi)\,C_s(\phi)\,ds
\end{align}
for some \textit{non-negative} function $f(s)$.
%This property (ii) means
%\begin{align}\label{reflection property}
%-\sum_{x\in\Lambda_+,y\in\Lambda_-}\phi^*(x)\Box(x,y)\phi(y)\in\mathcal{\bar P}
%\end{align}
%where $\mathcal{\bar P}$ is the closure of $\mathcal{P}$ with respect to the convergence defined by eq.\eqref{sup norm}:
%defined in \eqref{sup norm}:\\
%\begin{align}
%\mathcal{\bar P}= \{\mathcal{F}\in\mathcal{A}\,|\, \exists \{\mathcal{F}_k\}_{k=1}^{\infty}\subset\mathcal{P}\,;\,
%\mathcal{F}=\lim_{k\to\infty}\mathcal{F}_k\}.
%\end{align}
In this bosonic system, while (i) holds true, 
the property (ii) breaks down, %(i.e. \eqref{reflection property} does not hold) 
as will be shown below.

To show this,
we will derive the spectral representation of the kernel $\Box(x,y)$. %By the Ginsparg Wilson relation,
First, the Fourier transformation $\Box(p)$ is given by:
\begin{align}\label{laplacian kernel}
\Box(p)
%&:=\Box {\rm e}^{ipx}/{\rm e}^{ipx}\no\\
&=1-\frac{1-\sum_\mu(1- \cos p_\mu)}{\sqrt{\sum_\mu \sin^2 p_\mu+\big[1-\sum_\mu(1- \cos p_\mu)\big]^2}}\no\\
&=1+\frac{b(\bvec p)-\cos p_0}{\sqrt{a(\bvec p)-2b(\bvec p)\cos p_0}},
\end{align}
where 
\begin{align}
a(\bvec p)&=1+\sum_j \sin^2 p_j+b(\bvec p)^2,\\
b(\bvec p)&=\sum_j(1- \cos p_j).
\end{align}
From this formula we obtain the following three-space representation of the kernel $\Box(x,y)$,
\begin{align}
\Box(x,y)\Big|_{x_0\not= y_0}=\int_{[-\pi,\pi]^3} \frac{d^3\bvec p}{(2\pi)^3}\,{\rm e}^{i\bvec p\cdot (\bvec x-\bvec y)}I(x_0-y_0;\bvec p),
%\underbrace{
\end{align}
where we have defined
\begin{align}
I(x_0-y_0;\bvec p)=\int_{-\pi}^{\pi} \frac{dp_0}{2\pi}\,
{\rm e}^{ip_0(x_0-y_0)}\frac{b(\bvec p)-\cos p_0}{\sqrt{a(\bvec p)-2b(\bvec p)\cos p_0}}.
%}_{=:I(x_0-y_0)}
\end{align}
\begin{figure}[b]
\begin{center}
\includegraphics[scale=0.68,trim=230 240 200 0,angle=-90]{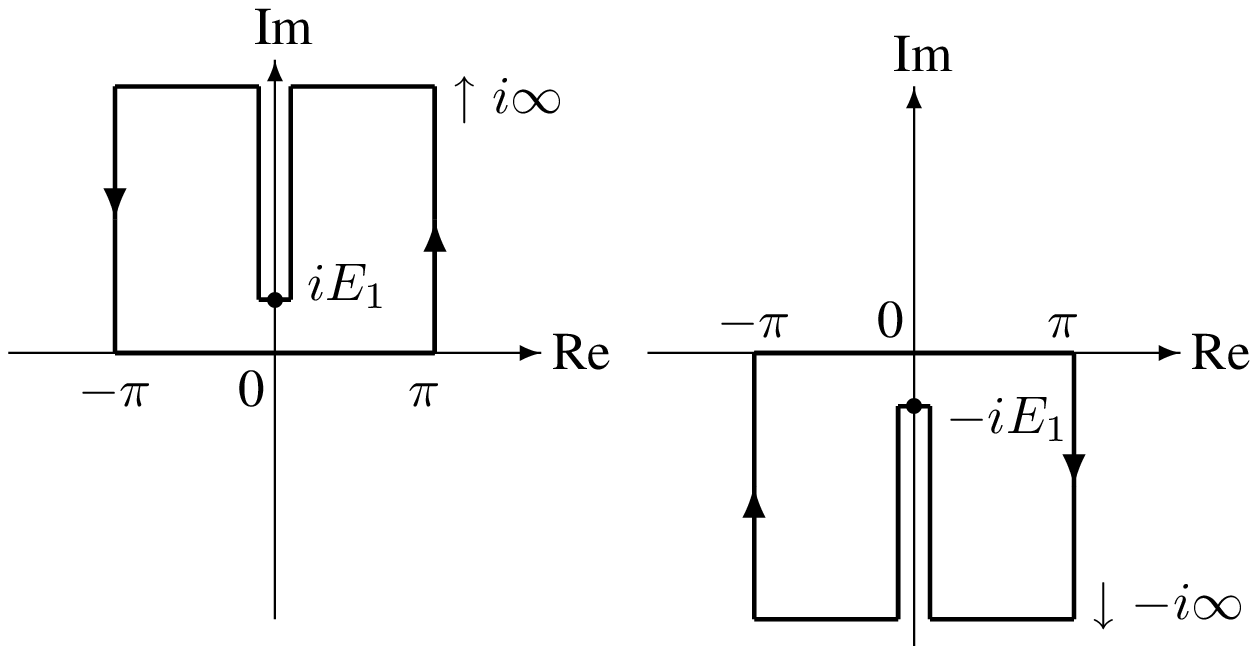}
\caption{Complex integration contours}
\label{fig:p0-integration-contour}
\end{center}
\end{figure}

We can transform $I(x_0;\bvec p)$ into a spectral representation 
by applying Cauchy's integration theorem to the contour in the complex $p_0$ plane drawn in the FIG.1.
In the FIG.1, the original integration contour for $p_0$  is the interval $[-\pi,\pi]$ on the real axis,
and $E_1$ is defined as the positive solution of 
\begin{align}
2b(\bvec p)\cosh E_1-a(\bvec p)=0.
\end{align}
In the case of $x_0>0$, we use the left contour and, in the case of $x_0<0$, we use the right. Note that the contributions coming 
from integrations along the edge with infinite (or minus infinite) imaginary part
 (upper edge in the case of $x_0>0$, lower in the case of $x_0<0$) vanishes.
Furthermore, because of the periodicity of the integrand, the contributions coming
 from integration along the edge whose real part is $\pi$ (the right edge of the contour) and from integration along the edge $-\pi$ (left edge
 of the contour) cancel each other. In this way, the original $p_0$ integration on the interval $[-\pi,\pi]$ can be expressed as 
 an integration on the interval $[iE_1,i\infty]$ (or $[-iE_1,-i\infty]$)
 along the imaginary axis: 
\begin{align}
I(x_0;\bvec p)\Big|_{x_0\not= 0}&=\int_{-\pi}^{\pi} \frac{dp_0}{2\pi}\,{\rm e}^{ip_0x_0}\frac{b(\bvec p)-\cos p_0}{\sqrt{a(\bvec p)-2b(\bvec p)\cos p_0}}\no\\
%&=-\int^{E_1}_\infty \frac{idE_1}{2\pi}e^{-E_1x_0}\frac{b(\bvec p)-\cosh E_1}{i\sqrt{2b(\bvec p)\cosh E_1-a(\bvec p)}}
%-\int_{E_1}^\infty \frac{idE_1}{2\pi}e^{-E_1x_0}\frac{b(\bvec p)-\cosh E_1}{-i\sqrt{2b(\bvec p)\cosh E_1-a(\bvec p)}}\no\\
&=\int_{E_1}^\infty \frac{dE}{\pi}\,{\rm e}^{-E|x_0|}\frac{b(\bvec p)-\cosh E}{\sqrt{2b(\bvec p)\cosh E-a(\bvec p)}}.
\end{align}
Since the kernel of the operator on the finite lattice $\Box_\Lambda(x,y)$ is defined as \eqref{box on lambda},
%\begin{align}
%D(x,y)=\sum_{n\in\Integer^d}D_\infty(x+2nL,y),
%\end{align} 
it is straightforward to derive the following spectral representation of $\Box_\Lambda$,
\begin{align}
&\Box_{\Lambda}(x,y)\Big|_{x_0>y_0}\no\\
&=\frac{1}{(2L)^3}\sum_{\bvec p}{\rm e}^{i\bvec p(\bvec x-\bvec y)}\sum_{n_0\in\Integer}I(x_0+2n_0L-y_0;\bvec p)\no\\
&=\frac{1}{(2L)^3}\sum_{\bvec p}{\rm e}^{i\bvec p(\bvec x-\bvec y)}\int _{E_1}^\infty
\frac{dE}{\pi}\frac{1}{1-{\rm e}^{-2EL}}\times \no\\
&\qquad\times\left[{\rm e}^{-E(x_0-y_0)}+{\rm e}^{-E(2L-x_0+y_0)} \right]
\frac{b(\bvec p)-\cos p_0}{\sqrt{a(\bvec p)-2b(\bvec p)\cos p_0}}.\label{spectral representation of DdaggerD}
\end{align}
The second term represents a finite volume effect.
 
 Now we can see the property \eqref{reflection property} does not hold. From eq.\eqref
 {spectral representation of DdaggerD}, we obtain
% Fisrt, note that $f(x_0-y_0)\ge 0$ for all $E$, $x_0$, and $y_0$. Second, 
 \begin{align}\label{spec rep}
&\quad-\sum_{x\in\Lambda_+,y\in\Lambda_-}\phi^*(x) \Box_{\Lambda}(x,y)\phi(y)\no\\
&=\frac{1}{(2L)^3}\sum_{\bvec p}\int_{E_1}^\infty \frac{dE}{\pi}\frac{1}{1-{\rm e}^{-2EL}}\frac{\cosh E-b(\bvec p)}{\sqrt{2b(\bvec p)\cosh E-a(\bvec p)}}\times\no\\
&\quad\times\sum_{x\in\Lambda_+,y\in\Lambda_-}\Big\{{\rm e}^{-E(x_0-y_0)}{\rm e}^{i\bvec p\cdot (\bvec x-\bvec y)}\phi^*(x)\phi(y)+\no\\
&\qquad\qquad\qquad+{\rm e}^{-2EL}{\rm e}^{+E(x_0-y_0)}{\rm e}^{i\bvec p\cdot (\bvec x-\bvec y)}\phi^*(x)\phi(y)\Big\}
\no\\
&=\frac{1}{(2L)^3}\sum_{\bvec p}\int_{E_1}^\infty \frac{dE}{\pi}\frac{1}{1-{\rm e}^{-2EL}}\frac{\cosh E-b(\bvec p)}{\sqrt{2b(\bvec p)\cosh E-a(\bvec p)}}\times\no\\
&\quad\times\Big\{\theta\big(C_{E,\bvec p}\big)(\phi)\big(C_{E,\bvec p}\big)(\phi)+\no\\
&\qquad+\theta\big(e^{-EL}C_{-E,\bvec p}\big)(\phi)\big(e^{-EL}C_{-E,\bvec p}\big)(\phi)\Big\},
\end{align}
where we define
\begin{align}
C_{E,\bvec p}(\phi)=\sum_{x\in\Lambda_+}{\rm e}^{-Ex_0}{\rm e}^{i\bvec p\cdot\bvec x}\phi^*(x)\in{\cal A_+}.
\end{align}
In this case, $(E,\bvec p)$ plays a role of the parameter $s$ in \eqref{reflection property}.
For the condition \eqref{reflection property} to be satisfied, the coefficient factor $\cosh E-b(\bvec p)$ should be non-negative 
for any $(E,\bvec p)$ satisfying $E_1\le E$, but this is \textit{not} the case.
In fact, $\cosh E-b(\bvec p)$ can become \textit{both} positive and negative in general, %depending on the spacial momenta $\bvec p$,
which prevents us from proving the reflection positivity.

\subsection{K\"{a}ll\'{e}n-Lehmann representation of the free overlap boson propagator}
In the previous section, we have shown that the standard way of proving the reflection positivity does not work for the overlap boson. In this section, we will investigate the 
K\"{a}ll\'{e}n-Lehmann representation of the propagator.
Here, we will observe that the spectral density function, which is expected to be 
positive (non-negative) for unitary quantum theories, is not a positive function. This implies
that the overlap boson system has pathological spectrum of
energy momentum operators.

The spectral density $\rho(E,\bvec p)$ is geven in the
the Euclidean version of 
K\"{a}ll\'{e}n-Lehmann representation by 
\begin{align}\label{def of spec dens}
\Delta_+(x,y)&:=\Expect{\phi(x)^*\phi(y)}\Big|_{x_0>y_0}\no\\
&=\int\frac{d^3\bvec p}{(2\pi)^3}\int_0^\infty\frac{dE}{\pi}\,{\rm e}^{-E(x_0-y_0)}{\rm e}^{-i\bvec p(\bvec x-\bvec y)}
\rho(E,\bvec p).
\end{align} 
In the present case,
one can explicitly estimate the propagator and the spectral density $\rho(E,\bvec p)$ by
using the Fourier transformation,
\begin{align}
\Delta_+(x,y)&=\frac{1}{\Box}(x,y)\Big|_{x_0>y_0}\no\\
&=\int\frac{d^4p}{(2\pi)^4}{\rm e}^{ip(x-y)}\frac{1}{\Box(p)}. %\no\\
%&=\int\frac{d^4p}{(2\pi)^4}{\rm e}^{ip(x-y)}\frac{1}{(D(p)+D(p)^\dagger)}
\end{align} 
The result is
\begin{align}
\rho(E,\bvec p)&=(\text{proportional to delta function})+\no\\
&+\frac{(\cosh E -b(\bvec p))\sqrt{2b(p)\cosh E-a(\bvec p)}}{\cosh^2E-a(\bvec p)+b(\bvec p)^2}\theta(E-E_1).\label{spectral density}
\end{align}
This formula is derived in Appendix \ref{app}. 
The second term, continuous spectrum, is not positive (non-negative)
because of the factor $\cosh E-b(\bvec p)$,
which is exactly the same one as appeared in \eqref{spec rep}.

The fact that the positivity of the spectral density breaks down
is an indirect but strong circumstantial evidence that
the overlap boson system does not define a quantum mechanical 
system with physically satisfactory energy-momentum spectrum. 
Further, it is very probable that the overlap boson system breaks the reflection
positivity condition. 
%, because this condition enables us to 
%construct a satisfactory quantum mechanical system corresponding 
%the lattice field theory. 

In fact, one can prove in mathematically rigorous manner 
\cite{Usui:2011} that a lattice theory satisfying the reflection positivity condition,
in addition to some technical assumptions which are satisfied by the overlap boson system,
must have the non-negative spectral density.
The proof goes as follows. For a detail, see \cite{Usui:2011}.   
Suppose that the reflection positivity is satisfied, and take the infinite volume limit $\Lambda\to\Integer^d$.
Then, from the reflection positivity condition, the Hilbert space of state vectors is constructed and for each $\mu=0,1,\dots,d-1$, 
momentum operator $P_\mu$
acting on this Hilbert space is defined as an infinitesimal self adjoint generator of translation in each direction. 
We denote $P_0=H$ and $\bvec P = (P_1,\dots, P_{d-1})$.
Since translations in different directions commute, $P_\mu$ and $P_\nu$ commute with each other if
$\mu\not=\nu$. Therefore $(H,\bvec p)$ possesses complete orthonormal set of simultaneous eigenvectors $\ket{E,\bvec p}$ with
$E\ge 0$, $\bvec p\in[-\pi,\pi)^{d-1}$.
Let $\ket 0$ be $\ket{0,\bvec 0}$.
Through momentum operators $(H,\bvec P)$,
field operator at $x=(x_0,\bvec x)\in\Integer^d$, $\hat{\phi}(x)$ is related to the field operator at the origin $\hat{\phi}(0)$ by
the relation
\begin{align}\label{trans of field op}
\hat{\phi}(x)={\rm e}^{-x_0 H}{\rm e}^{-i\bvec P\cdot \bvec x}\,\hat{\phi}(0)\,{\rm e}^{x_0 H}{\rm e}^{i\bvec P\cdot x}.
\end{align}
The two point function is expressed in terms of these ingredients of Hilbert space, 
\begin{align}
\Expect {\phi(x)^*\phi(y)}=\bra 0 \hat{\phi}(x)^\dagger\hat{\phi}(y) \ket 0.
\end{align}
By the translational invariance and \eqref{trans of field op}, and by inserting the identity
in the form of
\begin{align}
1=\int_{[0,\infty)\times [-\pi,\pi)^{d-1}} dE\,d^3\bvec p\, \ket{E,\bvec p}\bra{E,\bvec p},
\end{align}
 we obtain 
\begin{align}\label{formal KL}
\Delta_+(x,y)&=\bra 0 \hat{\phi}(x)^\dagger\hat{\phi}(y) \ket 0\Big|_{x_0>y_0}\no\\
&=\bra 0 \hat{\phi}(x-y)^\dagger\hat{\phi}(0) \ket 0\no\\
&=\bra 0 \hat{\phi}(0)^\dagger{\rm e}^{-(x_0-y_0) H}{\rm e}^{-i\bvec P\cdot (\bvec x-\bvec y)}\hat{\phi}(0) \ket 0\no\\
&=\int_{[0,\infty)\times [-\pi,\pi)^{d-1}} d |\bra 0 \hat{\phi}(0)\ket{E,\bvec p}|^2 \no\\
&\qquad \qquad \qquad{\rm e}^{-E(x_0-y_0)}{\rm e}^{-i\bvec p(\bvec x-\bvec y)},
\end{align}
where the integration measure is formally given by
\begin{align}\label{meas}
d |\bra 0 \hat{\phi}(0)\ket{E,\bvec p}|^2=d^3\bvec p \,dE  \,|\bra 0 \hat{\phi}(0)\ket{E,\bvec p}|^2.
\end{align}
Even though the above computation is rather formal, the measure \eqref{meas}
has mathematically rigorous meaning as a $d$-dimensional Borel measure supported on $[0,\infty)\times [-\pi,\pi)^{d-1}$.
In fact, one can derive \eqref{formal KL} in mathematically rigorous manner. 
It is well known in measure theory that any measure $\mu$ on $\Real^d$ can be uniquely decomposed into two parts 
\begin{align}\label{L decomp}
\mu=\mu_{\text{s}}+\mu_{\text{abs}}
\end{align}
where $\mu_{\text{s}}$ is singular and $\mu_{\text{abs}}$ is absolutely continuous
with respect to the Lebesgue measure on $\Real^d$ (Lebesgue decomposition theorem).
In more ordinary expression in physics, this theorem states that if we write 
\begin{align}
d |\bra 0 \hat{\phi}(0)\ket{E,\bvec p}|^2=\frac{d^3\bvec p}{(2\pi)^3} \,\frac{dE}{\pi} \, \rho(E,\bvec p),
\end{align}
$\rho(E,\bvec p)$ can be written corresponding to \eqref{L decomp} as
\begin{align}\label{rho decomp}
\rho(E,\bvec p)=(\text{singular part})+\rho_c(E,\bvec p),
\end{align}
where $\rho_c(E, \bvec p)$ is \textit{non-negative} integrable function.
This is not the case for the overlap boson.

\subsection{Estimation of the violation in the momentum space}
From the explicit form of the spectral density \eqref{spectral density}, 
we can find where in the Brillouin zone the reflection positivity is violated. %that if we restrict the summation over the spacial momenta $\bvec p$, we can avoid 
%the violation of the reflection positivity. 
One notes that there is the region in the spacial Brillouin zone where the spectral density $\rho(E,\bvec p)$
can not become negative. Let us call this region $\mathcal{S}$. The region $\mathcal{S}$ is characterized by the condition 
that the negative value of $\cosh E-b(\bvec p)$ be avoided. The necessary and sufficient condition
on spacial momenta $\bvec p$ to avoid negative 
$\cosh E-b(\bvec p)$ is that
\begin{align}
\cosh E -b(\bvec p)\ge 0,\qquad \forall E \ge E_1,
 \end{align}
which is equivalent to the condition $\cosh E_1\ge b(\bvec p)$
\begin{align}\label{safe condition}
 \text{i.e.} \quad\frac{1+\sum_k\sin^2 p_k-(\sum_{k}(1-\cos p_k))^2}{2\sum_{k}(1-\cos p_k)}\ge 0,
\end{align}
or, equivalently,
\begin{align}
1+\sum_{k}\sin^2 p_k - b(\bvec p)^2 \ge 0.
\end{align}
Then, define
\begin{align}
{\cal S}=\left\{\bvec p \in [-\pi,\pi)^{d-1} \,:\, 1+\sum_{k}\sin^2 p_k-b(\bvec p)^2 \ge 0 \right\},
\end{align}
where $d$ is the spacetime dimension.

Now let us estimate the size of $\mathcal{S}$ %. In other words, 
to investigate whether we can ignore the violation of the reflection 
positivity or not. % in practical point of view by giving some simple numerical examples.
In the case of $d=4$, there are three spacial momentum components.
First, we consider the case in which $p_1=p_2=p_3=:p$.
In this direction, the safe momentum region has the extent
\begin{align}
-1.84\lesssim p \lesssim 1.84.
\end{align} 
Second, we consider another direction $p_1=p,\,p_2=p_3=0$. In this case, in the safe region $\mathcal{S}$, $p$
is restricted by
%\footnote{This restriction value of spacial momenta $\bvec p$ is exactly the same as for the two dimensional case.
%In the case of $d=2$, the spacial momenta has only one component and ${\cal S}$ is given as
%\begin{align}
%{\cal S}=\{-1.95\le p \le 1.95 \}.
%\end{align} 
%}
\begin{align}
-2.23\lesssim p \lesssim 2.23.
\end{align} 
These regions are a little bit lager than $[-\pi/2,\pi/2]^{d-1}$. 
%This may mean that
%there is little problem of unitarity at least
%in practice, if we consider only the phenomena
%whose typical energy scale is much smaller than the half of the natural cut off determined by the lattice spacing. 

 When the spacial momenta $\bvec p$ does not belong to $\mathcal{S}$, the spectral density $\rho(E,\bvec p)$ has to become
negative on the energy interval $E_1\le E < E_c$, where $E_1$ and $E_c$ are determined by
% These energies are characterized by the relations
\begin{align}
\cosh E_1=\frac{a(\bvec p)}{2b(\bvec p)},\qquad \cosh E_c=\bvec b(\bvec p),
\end{align} 
since $\rho(E,\bvec p)< 0$ is equivalent to $a(\bvec p)/2b(\bvec p)\le \cosh E < b(\bvec p)$ when $\bvec p\not\in\mathcal{S}$.
We will numerically estimate $E_1$ and $E_c$, the lower and upper bound of the energy interval on which 
the spectral density become negative. 
%which may indicate
%he energy scale of positivity violation, in the case where $\bvec p$ does \textit{not} belong to $\mathcal{S}$. 
%On the other hand, if the spacial momenta $\bvec p$ doesn't belong to ${\cal S}$,  is estimated depending on the value of spacial momenta. 
For instance, if $d=4$, these energy values are computed as shown in the following table:
\begin{center}
\begin{tabular}{|c||c|c|c|c|c|}
\hline
$\bvec p$ & $b(\bvec p)$ & $a(\bvec p)$ & $a(\bvec p)/2b(\bvec p)$ & $E_1$ & $E_c$ \\
\hline
$(\pi,\pi,\pi)$ & $6$ & $37$ & $37/12$ & $1.79\dots$ & $2.48\dots$\\
\hline
$(\pi,\pi,0)$ & $4$ & $17$ & $17/8$ & $1.39\dots$ & $2.06\dots$\\
\hline
$(\pi,0,0)$ & $2$ & $5$ & $5/4$ & $0.69\dots$ & $1.32\dots$\\
\hline
\end{tabular}
\end{center}
%
%\begin{align}
% \bvec p &=(\pi,\pi,\pi)\,\Longrightarrow \,E_1=\cosh^{-1}\frac{37}{12}\approx 1.79 ,\quad E_c=\cosh^{-1} 6\approx2.48,\\
% \bvec p &=(\pi,\pi,0)\,\Longrightarrow \,E_1=\cosh^{-1}\frac{17}{8}\approx 1.39 ,\quad E_c=\cosh^{-1} 4\approx2.06,\\
% \bvec p &=(\pi,0,0)\,\Longrightarrow \,E_1=\cosh^{-1}\frac{5}{4}\approx 0.69,\quad E_c=\cosh^{-1} 6\approx1.32.
%\end{align}
%
Whether these values are large enough or not should depend on the physics one wants to see
through the overlap boson.

\section{Refletion positivity of lattice Wess-Zumino model}
\label{sec:RF-lattice-WZ-model}

To remedy the violation of the reflection positivity, 
one may adopt the simpler nearest-neighbor action for the boson fields, 
$\phi$ and $F$ as follows\footnote{
Here, we have changed the sign convention of the fermionic action by introducing new Majorana field $\chi'=i\chi$.
Of course this does not change any physical results. It is simply because 
this convention has been used in the proof of the reflection positivity for the overlap fermions
in our previous work \cite{Kikukawa:2010gq}.
}:
\begin{eqnarray}\label{positive action}
   S_0^\prime=\sum_x\Bigl\{ \, 
   -{1\over2}\chi^TCD \chi +\phi^*  (-\partial_\mu^* \partial_\mu )   \phi+F^*F   \, \Bigr\} . 
%\nonumber\\
%   &&\qquad\qquad\qquad\quad
%   -{1\over a}X^TCX
%   -{2\over a}(\mathcal{F}\Phi+\mathcal{F}^*\Phi^*)
%   \, \Bigr\},
\end{eqnarray}
This action still possesses three types of $\U(1)$
symmetry, Eq.~(\ref{twoxfifteen}), (\ref{twoxseven}) and 
\begin{eqnarray}
%   &&\delta_\alpha\chi=0,\qquad \delta_\alpha X=0,
%\nonumber\\
   &&\delta_\alpha\phi  =+i\alpha \phi , %+i\alpha\Phi,
\nonumber\\
   &&\delta_\alpha F=+i\alpha F , 
\label{twoxsixteen-2}
\end{eqnarray}
instead of Eq.~(\ref{twoxsixteen}). 

In this formulation of the chiral multiplet,  the action of  the lattice ${\cal N}=1$ Wess-Zumino model may be given as follows:
\begin{align}
\label{eq:WZ-model-action-nearest-neighbor}
   S&=\sum_x\Bigl\{-{1\over2}\chi^TCD\chi
   +\phi^*(-\partial_\mu^* \partial_\mu )\phi+F^*F    +X^TCX
\nonumber\\
   &\quad
   -g\tilde \chi^T C \phi P_+ \tilde \chi 
   -g^*\tilde \chi^T C\phi^*P_-\tilde \chi +g F \phi^2+g^* F^*\phi^{*2}
   \, \, \Bigr\},
\end{align}
where $X(x)$ is an auxiliary Majorana fermion field and  $\tilde \chi(x) = \chi(x) + X(x)$. 
Then one may define the $\U(1)_R$ symmetry as follows: 
\begin{eqnarray}
   &&\delta_\alpha\chi=+i\alpha\gamma_5(1-{1\over2}D)\chi , 
%   +i\alpha\gamma_5X,\qquad
%   \delta_\alpha X=+i\alpha\gamma_5{1\over2}aD\chi,
\nonumber\\
   &&\delta_\alpha\phi
   =-2i\alpha\phi, 
\nonumber\\
%   &&\delta_\alpha\Phi=-3i\alpha\Phi
%   +i\alpha\{{1\over2}aD_2\phi+{1\over2}aF^*\},
%\nonumber\\
   &&\delta_\alpha F=+4i\alpha F
%   +i\alpha\mathcal{F},
%\nonumber\\
%   &&\delta_\alpha\mathcal{F}=+3i\alpha\mathcal{F}
%   +i\alpha\{{1\over2}aD_2F+{1\over2}aD_1^2\phi^*\}.
\label{twoxtwentythree}
\end{eqnarray}

The reflection positivity is now satisfied in this formulation of the Wess-Zumino model. 
%both in the fermionic and bosonic parts 
%of the free action, Eq.~(\ref{positive action}), 
%and also  in the action of the Wess-Zumino model, Eq.~(\ref{eq:WZ-model-action-nearest-neighbor}),  including the interaction terms. 
%The interaction terms are all strictly 
%local and does not cause any problem in proving the reflection positivity with respect to the link reflection.  
The $\theta$-reflection is defined for the bosonic fields $\chi,F$ 
in the same way as in the generic case \eqref{theta for boson}, 
\begin{align}
\theta\phi(x)=\phi(\theta x) \\
\theta F(x)=F(\theta x)
\end{align}
and for the fermionic fields $\chi,X$ as in \eqref{theta for fermion},
\begin{align}
(\theta\bar\chi)(x)&=\gamma_0\chi(\theta x),  \qquad (\theta\chi)(x)=\bar\chi(\theta x)\gamma_0,\\
(\theta\bar X)(x)&=\gamma_0 X(\theta x),  \qquad (\theta X)(x)=\bar X(\theta x)\gamma_0.
\end{align}
Note that this definition of $\theta$ reflection does not contradict to the Majorana conditions $\bar\chi=\chi^T C$
and $\bar X=X^T C$.
Our field algebra ${\cal A}$ here is that of the polynomial algebra of fermionic fields whose coefficients are
the well-behaved functions of the bosonic fields. We extend $\theta$ operation to whole algebra ${\cal A}$,
by the relations \eqref{product} and \eqref{linear combi}.

%For any observable ${\cal F}$ of the form ${\cal F}(\chi, X, \phi, F)=f(\phi, F)M(\chi,X)$ with $f$ being some well-behaved 
%function of $\{\phi(x),F(x)\}_{x\in\Lambda}$ and $M(\chi,X)$ some monomial of $\{\chi_\alpha(x),\bar\chi_\alpha(x),
%X_\alpha(x),\bar X_\alpha(x)\}_{x\in\Lambda}$, we define 
%\begin{align}
%\theta(\mathcal{F})(\chi, X, \phi, F)=f^*(\theta\phi,\theta F)M^\dagger(\theta\chi,\theta X),
%\end{align}
%where $*$ means complex $M^\dagger$ means the monomial whose order of the Grassmann product
%is reversed in the original $M$. We extend the $\theta$ operation for arbitrary observables by anti-linearity.

%This action does violate the vacuum energy cancelation but preserves the reflection positivity as well
%as the $U(1)_R$ symmetry. 

%Let us show the reflection positivity of this Wess-Zumino model.
%By slightly modifying the previous discussion, 
%we can see that, even in the case where there are both boson, and fermion 
%it is sufficient to see that action can be rewritten in the form of

To prove the reflection positivity of the Wess-Zumino model, 
it is sufficient to show that the action \eqref{eq:WZ-model-action-nearest-neighbor} can be rewritten in the form of
\eqref{decomposition of the action}
\begin{align}\label{positive form}
-S=B + \theta(B)+\sum_s\theta(C_s)C_s,  \quad B,C_s\in\mathcal{A}_+ , 
\end{align}
%or,  in the form of 
\cite{Kikukawa:2010gq}. 
Let us first consider the free part of \eqref{eq:WZ-model-action-nearest-neighbor}.  
The first term in \eqref{eq:WZ-model-action-nearest-neighbor}, 
the overlap Majorana fermion,  can be written in the form of \eqref{decomposition of the action} as is shown in ref \cite{Kikukawa:2010gq}. 
On the other hand, the second term can  be written in the form of \eqref{positive form}, as is well-known. 
Furthermore, the third and fourth terms in 
\eqref{eq:WZ-model-action-nearest-neighbor} is $\theta$ reflection of themselves and do not contain any
`time hopping terms'. Therefore, these terms can be written in the form of
\begin{align}
&\sum_{x_\in\Lambda}\Big\{ F^*F +X^TCX \Big\}\no\\
&=\sum_{x\in\Lambda_+}\Big\{ F^*F+X^TCX \Big\} + \theta\Big[\sum_{x\in\Lambda_+}\Big\{
 F^*F+X^TCX \Big\}\Big].
\end{align}
The rest of the terms in \eqref{eq:WZ-model-action-nearest-neighbor}  are interaction terms, 
\begin{align}
S_{\text{int}}&:=\sum_x\Bigl\{
  - g\tilde \chi^T C \phi P_+ \tilde \chi 
   -g^*\tilde \chi^T C\phi^*P_-\tilde \chi +\no\\
   &\qquad\qquad+g F \phi^2+g^* F^*\phi^{*2}
   \, \, \Bigr\},
\end{align}
%The interaction terms are all strictly 
%local and does not cause any problem in proving the reflection positivity with respect to the link reflection.  
which are all strictly local. 
They are equal to theta-reflection of themselves again, and 
do not contain any nonlocal `time hopping' terms either. %, and . 
This means that $S_\text{int}$ can also be written as
\begin{align}
S_{\text{int}}=B+\theta(B)
\end{align}
with
\begin{align}
B=&\sum_{x\in\Lambda_+}\Bigl\{
   -g\tilde \chi^T C \phi P_+ \tilde \chi 
   -g^*\tilde \chi^T C\phi^*P_-\tilde \chi +\no\\
   &\qquad\qquad +g F \phi^2+g^* F^*\phi^{*2} 
   \, \, \Bigr\},
\end{align}
which obviously belongs to ${\cal A}_+$. Therefore,  one concludes that this lattice Wess-Zumino model
satisfies the reflection positivity condition.

\section{Discussion} 
\label{sec:discussion} 

Preserving R symmetry exactly is a useful way in formulating supersymmetric field theories on the lattice.
This point has been emphasized by Elliot, Giedt and Moore \cite{Elliott:2008jp} 
in their formulation of four-dimensional ${\cal N}=4$ super Yang-Mills theory. 
The discrete R symmetry in the two-dimensional ${\cal N}=2$
Wess-Zumino model \cite{Kikukawa:2002as} has played an important role in the numerical study 
of the correspondence to  ${\cal N}=2$ conformal field theories\cite{Kawai:2010yj}. 

In formulating the exact R symmetry on the lattice, however, there is a freedom in the choice of the bosonic part of the action. 
When one can preserve some part of the extended supersymmetries 
in the theories with ${\cal N} \ge 2$ \cite{Kikukawa:2002as, Kikukawa:2008xw}, 
it seems useful to adopt the bosonic actions to preserve the supersymmetries, 
although one should take into care a possible effect of the violation of the reflection positivity. 
But,  for the theories of ${\cal N} = 1$, it seems difficult to preserve 
the supersymmetry in general\cite{Kato:2008sp}, and 
the free limit supersymmetry does not necessarily help in taking the supersymmetric limit in the interacting models.
In such situations, thought, if one can preserve the fundamental requirement of the reflection positivity condition, it may serve as a possible guideline to choose a bosonic action.

%one may impose the reflection positivity condition as   

It would be interesting to examine further the inter-relation among
the reflection positivity, the vacuum energy cancellation(the exact supersymmetry) and 
the exact U(1)$_R$ symmetry of free chiral multiplet on the lattice. 
If one adopts the Majorana Wilson fermion for the fermionic component of the chiral multiplet,  
one can show that the bosonic part of the supersymmetric action now fulfills the reflection positivity condition.  
In this case, the U(1)$_R$ symmetry is not manifest.
But, through the block spin transformation, it is recovered in the fixed point action \cite{Ginsparg:1981bj}.
%Also, by adjusting the parameters in the block-spin kernels and the normalization factors, 
In this course of the renormalization group transformatons, 
it seems possible to maintain the vacuum energy cancellation
by adjusting the parameters in the block-spin kernels and the normalization factors. 
Then,  if  the reflection positivity could also be maintained through the block-spin transformation, 
all the three conditions could be fulfills in the fixed point 
approach\cite{Hasenfratz:1998ri, Hasenfratz:1998jp, Bell-Wilson:1975, Wiese:1993cb,So:1998ya}. 
%Work to clarify this point is in progress. 

\section*{Acknowledgements}
K.U. would like to thank Tsutomu T. Yanagida for continuous encouragement.
He also thanks Kenji Maeda for valuable discussions.
K.U. is supported by Global COE Program ``the Physical Science Frontier", MEXT, Japan.
This work was supported by World Premier International Center Initiative (WPI Program), MEXT, Japan.
%Y.K. would like to thank S.~Hashimoto and H.~Fukaya for discussions.
Y.K. is supported in part by Grant-in-Aid for Scientific Research No.~21540258, ~21105503.

%%%%%%%%%%%%%% APPENDIX %%%%%%%%%%%%%%%%%%%%%%
\appendix
\section{Spectral density $\rho$}\label{app}
In this appendix, we derive the formula \eqref{spectral density} for $d=4$. 
To find the explicit form of the spectral density $\rho$, we 
express the propagator by its Fourier transformation :
\begin{align}\label{propagator}
\Delta_+(x,y)&=\int\frac{d^4p}{(2\pi)^4}{\rm e}^{ip(x-y)}\frac{1}{\Box}(p_0,\bvec p)\no\\
&=\int\frac{d^3\bvec p}{(2\pi)^3}{\rm e}^{i\bvec p (\bvec x-\bvec y)}\int\frac{dp_0}{2\pi}{\rm e}^{ip_0 (x_0-y_0)}
\frac{1}{\Box}(p_0,\bvec p).
\end{align}
In the following analysis, we will apply Cauchy's theorem to the $p_0$ integration
\begin{align}
\int\frac{dp_0}{2\pi}{\rm e}^{ip_0 (x_0-y_0)}
\frac{1}{\Box}(p_0,\bvec p).
\end{align}
Note that though the integration \eqref{propagator}
is well defined, the integrand function of $\bvec p$ in \eqref{propagator}:
\begin{align}\label{p_0 func}
\bvec p \mapsto \int\frac{dp_0}{2\pi}{\rm e}^{ip_0 (x_0-y_0)}
\frac{1}{\Box}(p_0,\bvec p)
\end{align}
is not defined at $\bvec p =0$, because at $\bvec p=0$ this $p_0$ integration does not exist.
Since the value of the integrand on a set of zero measure $\{ (p_0,\bvec p) \,:\, \bvec p =0 \}$ 
does not contribute the integration, we may assume $\bvec p\not=0$ (mod $2\pi$) in \eqref{p_0 func} when applying
 Cauchy's theorem.

Define a function $f(z)$ as
\begin{align}
f(z)=\frac{1}{\Box}(p_0=z,\bvec p).%\qquad \mathrm{Re}\,z\in(-\pi,\pi].%=\frac{1}{D+D^\dagger}(z,\bvec p),\qquad -\pi<{\rm Re}\,z\le \pi.
\end{align}
From eq.\eqref{laplacian kernel}, one finds
\begin{align}
f(z)=\frac{\sqrt{a-2b\cos z}}{\sqrt{a-2b\cos z}+b-\cos z},
\end{align}
where, for notational simplicity, we have written $a=a(\bvec p)$ and $b=b(\bvec p)$. 
Here, we have to clarify the meaning of the square root of complex variables.
We define the square root of 
\begin{align}
z=|z| {\rm e}^{i\theta}, \quad\theta\in(-\pi,\pi),
\end{align}
as 
\begin{align}
\sqrt{z}:=\sqrt{|z|} {\rm e}^{i\theta/2},\quad \theta\in(-\pi,\pi).
\end{align}
Namely, we choose the branch where ${\rm Re}\,\sqrt{z}\ge 0$. 

We investigate the analytic structure of $f$.

First, since the square root of negative real number cannot be defined, 
$f$ cannot be defined where
\begin{align}\label{branch cut}
a-2b\cos z <0.
\end{align}
To find the more explicit condition which is equivalent to \eqref{branch cut},
put $z=x+iy\,(x,y\in\Real)$. Since $\cos(x+iy)=\cos x\cosh y-i\sin x\sinh y$, 
\eqref{branch cut} is equivalent to
\begin{align}
\cos x\cosh y-i\sin x\sinh y>\frac{a}{2b},
\end{align}
which holds true when and only when
\begin{align}\label{branch cut 2}
\cos x\cosh y>\frac{a}{2b}, \quad \sin x\sinh y=0.
\end{align}
The second condition is equivalent to 
\begin{align}
x=n\pi\,(n\in\Integer)\quad\text{or} \quad y=0, 
\end{align}
but the second choice $y=0$ is impossible because
in this case the first condition of \eqref{branch cut 2}
becomes
\begin{align}
\cos x > \frac{a}{2b} \ge 1,
\end{align}
which is never true for real $x$. Therefore, \eqref{branch cut 2}
is equivalent to $x=n\pi$, $n\in\Integer$, and
\begin{align}
\cos (n\pi) \cosh y = (-1)^n \cosh y >\frac{a}{2b},\quad (n\in\Integer).
\end{align}
Hence the condition \eqref{branch cut} occurs when and only when
\begin{align}
x=2n\pi\,(n\in\Integer),
\end{align}
and
\begin{align}
y<-E_1\quad \text{or}\quad E_1<y.
\end{align}

Next, we investigate the pole type singularity of $f$ which may appear where the
denominator 
\begin{align}
g(z):=\sqrt{a-2b\cos z}+b-\cos z
\end{align}
vanishes. To find the \textit{necessary} condition of $g(z)=0$, let us assume $g(z)=0$.
Then, by taking the square of the both sides of
\begin{align}
\sqrt{a-2b\cos z}=-b+\cos z,
\end{align}
one finds
\begin{align}
\cos^2 z=a-b^2.
\end{align}
Using the identity $\cos^2 z=(1+\cos 2z)/2$ and putting $z=x+iy \,(x,y\in\Real)$ again, one arrives at
\begin{align}\label{cos cosh}
\cos 2x\cosh 2y = 2a-2b^2-1
\end{align} 
and 
\begin{align}\label{sin sinh}
\sin 2x\sinh 2y =0.
\end{align}
Eq. \eqref{sin sinh} implies 
\begin{align}\label{cases}
y=0 \quad\text{or}\quad 2x=n\pi\,(n\in\Integer),
\end{align}
and we consider both cases respectively.

In the first case, $y=0$, \eqref{cos cosh} becomes
\begin{align}\label{cos 2x}
\cos 2x =2a-2b^2-1 \ge 1,
\end{align}
which is possible only when $2a-2b^2-1=1$, equivalently,
\begin{align}
\sum_{k=1}^3 \sin^2 p_k =0.
\end{align}
Therefore, this case $y=0$ occurs only when the spacial momentum satisfies
\begin{align}\label{cond for spacial p}
\bvec p=(m_1\pi,m_2\pi,m_3\pi),\quad m_1,m_2,m_3\in\Integer.
\end{align} 
As we noted at the beginning of this appendix, we may assume at least one of the
$m_k$'s ($k=1,2,3$) is odd (otherwise the $p_0$ integration is ill defined).
If \eqref{cond for spacial p} is satisfied, the right hand side of \eqref{cos 2x} becomes $1$ and 
\eqref{cos 2x} implies
\begin{align}
x=n\pi,\quad n\in\Integer.
\end{align}
But, this condition, $y=0$ and $x=n\pi\,(n\in\Integer)$, is not sufficient for $g(z)=0$. In fact, for $n\in\Integer$,
\begin{align}
g(n\pi)&=\sqrt{a-2(-1)^n b}+b-(-1)^n \no\\
 &=|b-(-1)^n|+b-(-1)^n \no\\
 &\ge2
\end{align} 
because
\begin{align}
b=b(\bvec p)=\sum_{k=1}^3(1-\cos p_k)\ge 2,
\end{align}
due to the fact that at least one of $\cos p_k$'s is equal to $-1$.

In the second case of \eqref{cases}, $2x=n\pi\,(n\in\Integer)$, \eqref{cos cosh}
becomes
\begin{align}
(-1)^n\cosh 2y = 2a-2b^2-1.
\end{align}
Then, this implies that $n$ is even and 
\begin{align}
\cosh 2y =2a-2b^2-1.
\end{align}
Define $E_0>0$ as 
\begin{align}
E_0 &= \frac{1}{2} \cosh^{-1}(2a-2b^2-1)\no\\
&=\cosh^{-1}\sqrt{a-b^2},
\end{align}
and we obtain as a necessary condition for $g(z)=0$, $z=n\pi\pm iE_0\,(n\in\Integer)$.
To find  a sufficient condition for $g(z)=0$, let us assume, 
Conversely, when $z=n\pi\pm iE_0$, then,
\begin{align}
g(n\pi\pm iE_0)&=\sqrt{a-2b(-1)^n\cosh E_0}+b-\cosh E_0\no\\
&=|b-(-1)^n\sqrt{a-b^2}|+ b-\sqrt{a-b^2}\no\\
&=\begin{cases}
0 \quad (\text{if}\;\; b-(-1)^n\sqrt{a-b^2}\le 0)\\
\\
2b\quad (\text{if}\;\;b-(-1)^n\sqrt{a-b^2}\ge 0 )
\end{cases}.
\end{align}
Hence, the necessary and sufficient condition for 
$g(n\pi\pm iE_0)=0$ is $n$ is even and
\begin{align}
b-\sqrt{a-b^2}\le 0,
\end{align}
which is equivalent to
\begin{align}\label{S cond}
1+\sum_{k=1}^3 \sin^2 p_k -b(\bvec p)^2\ge 0,
\end{align}
namely, $\bvec p \in\mathcal{S}$.

We now have found all the zeros of the function $g(z)$: 
\begin{align}
z=2n\pi \pm iE_0, \quad (n\in\Integer, \,\bvec p \in \mathcal{S}).
\end{align}

For a moment, let us assume spacial momentum $\bvec p$ satisfies
\begin{align}
b-\sqrt{a-b^2}< 0.
\end{align}
In this case, $z=z_n^\pm:=2n\pi \pm iE_0$ is
a simple pole of $f$, as will be seen. From the above argument, $f$ is 
analytic on 
\begin{align}
\Complex\setminus\bigcup_{n\in\Integer}\Big(\{z_n^\pm\}\cup\{2n\pi+iy\,;\,y<-E_1,\,E_1<y\}\Big).
\end{align}
Expand $g$ in Taylor series around $z_n^\pm$ :
\begin{align}
g(z)=g(z_n^\pm)+g'(z_n^\pm)(z-z_n^\pm)+\mathcal{O}\big({(z-z_n^\pm)}^2\big),
\end{align}
on $|z-z_n^\pm|<r$ for sufficiently small $r>0$, and we obtain
\begin{align}
f(z)&=\frac{\sqrt{a-2b\cos z}}{g(z_n^\pm)+g'(z_n^\pm)(z-z_n^\pm)+\mathcal{O}\big({(z-z_n^\pm)}^2\big)}\no\\
&=\frac{\sqrt{a-2b\cos z}}{g'(z_n^\pm)(z-z_n^\pm)+\mathcal{O}\big({(z-z_n^\pm)}^2\big)},
\end{align}
on $|z-z_n^\pm|<r$. Then, we find
\begin{align}
z\mapsto(z-z_n^\pm)f(z)=\frac{\sqrt{a-2b\cos z}}{g'(z_n^\pm)+\mathcal{O}\big(z-z_n^\pm\big)}
\end{align}
is analytic on $|z-z_n^\pm|<r$, and then $z=z_n^\pm$ are simple poles of $f$ with residues
\begin{align}\label{res}
\text{Res}(f,z_n^\pm)&=(z-z_n^\pm)f(z)\Big|_{z=z_n^\pm}\no\\
&=\frac{\sqrt{a-2b\cos (2n\pi \pm iE_0)}}{g'(2n\pi \pm iE_0)}\no\\
&=\pm\frac{a-2b\sqrt{a-b^2}}{i\sqrt{a-b^2}\sqrt{a-b^2-1}}.
\end{align}

Applying Cauchy's theorem on the contour drawn in FIG.2.,
we obtain, for $x_0>0$,
\begin{align}\label{Cauchy}
&\int_{[-\pi,\pi]}\frac{dp_0}{2\pi}\,{\rm e}^{ip_0 x_0}
\frac{1}{\Box}(p_0,\bvec p)\no\\
=&
\int_{[-\pi,\pi]}\frac{dz}{2\pi}\,{\rm e}^{izx_0}f(z)\no\\
=&
2\pi i \,\text{Res}(f,z_0)-\Bigg(\int_{i\infty+0}^{iE_1+0}+\int_{i\infty-0}^{iE_1-0}\Bigg)\frac{dz}{2\pi}\,{\rm e}^{iz x_0}f(z).
\end{align} 
Recalling our definition of the square root, one finds
\begin{align}
\sqrt{a-2b\cos(iE\pm 0)}=\pm i\sqrt{2b\cosh E-a}.
\end{align}
Then, the integrations of the second term in \eqref{Cauchy} is computed
by putting $z=iE\pm 0$ to become
\begin{align}\label{edge int}
&\Bigg(\int_{i\infty+0}^{iE_1+0}+\int_{i\infty-0}^{iE_1-0}\Bigg){\rm e}^{-E x_0}f(z)\no\\
=&\int_{E_1}^\infty\frac{i dE}{2\pi}{\rm e}^{-Ex_0}\Big(f(iE+0)+f(iE-0)\Big)\no\\
=-&\int_{E_1}^\infty\frac{dE}{\pi}{\rm e}^{-Ex_0}\frac{(b-\cosh E)\sqrt{2b\cosh E-a}}{\cosh^2 E-a+b^2}.
\end{align}
By substituting \eqref{res} and \eqref{edge int} into \eqref{Cauchy}, we arrive at
\begin{align}\label{answer of spec}
&\int_{[-\pi,\pi]}\frac{dp_0}{2\pi}\,{\rm e}^{ip_0 x_0}
\frac{1}{\Box}(p_0,\bvec p)\no\\
=&2\pi \frac{a-2b\sqrt{a-b^2}}{\sqrt{a-b^2}\sqrt{a-b^2-1}}\no\\
&\qquad+\int_{E_1}^\infty\frac{dE}{\pi}{\rm e}^{-Ex_0}\frac{(b-\cosh E)\sqrt{2b\cosh E-a}}{\cosh^2 E-a+b^2},
\end{align}
in agreement with \eqref{spectral density}.

Considering the case where spacial momentum $\bvec p$ satisfies
\begin{align}
b-\sqrt{a-b^2}\ge 0,
\end{align}
we find that there is no pole term and only the second term of \eqref{answer of spec} survives.
Note that, in the case of equality, even though $g(z)=0$, $f$ has no isolated pole.
In this case, the numerator of $f(z)$ also vanishes and $E_0=E_1$.

\begin{figure}[b]
\begin{center}
\includegraphics[scale=0.55,trim=100 200 100 250]{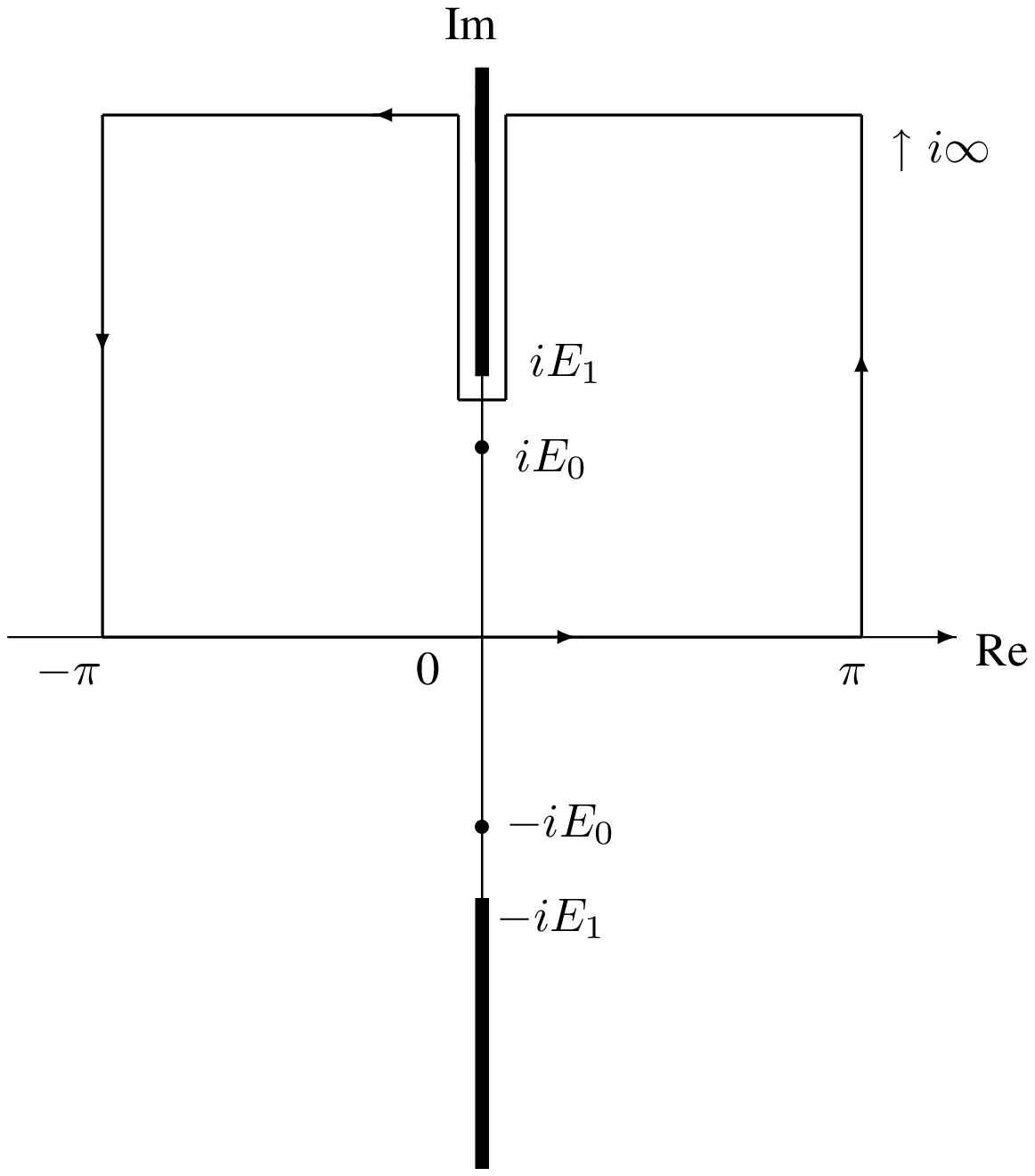} 
\caption{Integration contour of $f$.}
\label{fig:p0-integration-contour}
\end{center}
\end{figure}


\begin{thebibliography}{99}

%\cite{Wess:1973kz}
\bibitem{Wess:1973kz}
  J.~Wess and B.~Zumino,
  %``A Lagrangian Model Invariant Under Supergauge Transformations,''
  Phys.\ Lett.\  B {\bf 49}, 52 (1974).
  %%CITATION = PHLTA,B49,52;%%

%\cite{Neuberger:1997fp}
\bibitem{Neuberger:1997fp}
H.~Neuberger,
%``Exactly massless quarks on the lattice,''
Phys.\ Lett.\ B {\bf 417}, 141 (1998) .
%[arXiv:hep-lat/9707022].
%%CITATION = HEP-LAT 9707022;%%

%\cite{Neuberger:1998wv}
\bibitem{Neuberger:1998wv}
  H.~Neuberger,
  %``More about exactly massless quarks on the lattice,''
  Phys.\ Lett.\  B {\bf 427}, 353 (1998) . 
 % [arXiv:hep-lat/9801031].
  %%CITATION = PHLTA,B427,353;%%

%\cite{Kikukawa:1997qh}
\bibitem{Kikukawa:1997qh}
  Y.~Kikukawa and H.~Neuberger,
  %``Overlap in odd dimensions,''
  Nucl.\ Phys.\  B {\bf 513}, 735 (1998)
  [arXiv:hep-lat/9707016].
  %%CITATION = NUPHA,B513,735;%%

%\cite{Hasenfratz:1998ri}
%\bibitem{Hasenfratz:1998ri}
%P.~Hasenfratz, V.~Laliena and F.~Niedermayer,
%``The index theorem in QCD with a finite cut-off,''
%Phys.\ Lett.\ B {\bf 427}, 125 (1998) . 
%[arXiv:hep-lat/9801021].
%%CITATION = HEP-LAT 9801021;%%

%\cite{Hasenfratz:1998jp}
%\bibitem{Hasenfratz:1998jp}
%P.~Hasenfratz,
%``Lattice QCD without tuning, mixing and current renormalization,''
%Nucl.\ Phys.\ B {\bf 525}, 401 (1998) .
%[arXiv:hep-lat/9802007].
%%CITATION = HEP-LAT 9802007;%%

 %\cite{Karsten:1980wd}
\bibitem{Karsten:1980wd}
  L.~H.~Karsten and J.~Smit,
  %``Lattice Fermions: Species Doubling, Chiral Invariance, And The Triangle
  %Anomaly,''
  Nucl.\ Phys.\  B {\bf 183}, 103 (1981).
  %%CITATION = NUPHA,B183,103;%%

%\cite{Nielsen:1980rz}
\bibitem{Nielsen:1980rz}
  H.~B.~Nielsen and M.~Ninomiya,
  %``Absence Of Neutrinos On A Lattice. 1. Proof By Homotopy Theory,''
  Nucl.\ Phys.\  B {\bf 185}, 20 (1981)
  [Erratum-ibid.\  B {\bf 195}, 541 (1982)].
  %%CITATION = NUPHA,B185,20;%%
  
  %\cite{Nielsen:1981xu}
\bibitem{Nielsen:1981xu}
  H.~B.~Nielsen and M.~Ninomiya,
  %``Absence Of Neutrinos On A Lattice. 2. Intuitive Topological Proof,''
  Nucl.\ Phys.\  B {\bf 193}, 173 (1981).
  %%CITATION = NUPHA,B193,173;%%


%\cite{Ginsparg:1981bj}
\bibitem{Ginsparg:1981bj}
  P.~H.~Ginsparg and K.~G.~Wilson,
  %``A Remnant Of Chiral Symmetry On The Lattice,''
  Phys.\ Rev.\  D {\bf 25}, 2649 (1982).
  %%CITATION = PHRVA,D25,2649;%%


  
%\cite{Luscher:1998pqa}
\bibitem{Luscher:1998pqa}
  M.~Luscher,
  %``Exact chiral symmetry on the lattice and the Ginsparg-Wilson relation,''
  Phys.\ Lett.\  B {\bf 428}, 342 (1998) . 
%  [arXiv:hep-lat/9802011].
  %%CITATION = PHLTA,B428,342;%%



 %\cite{Aoyama:1998in}
\bibitem{Aoyama:1998in}
  T.~Aoyama and Y.~Kikukawa,
  %``Overlap formula for the chiral multiplet,''
  Phys.\ Rev.\  D {\bf 59}, 054507 (1999)
  [arXiv:hep-lat/9803016].
  %%CITATION = PHRVA,D59,054507;%%

%\cite{Fujikawa:2001ka}
\bibitem{Fujikawa:2001ka}
  K.~Fujikawa and M.~Ishibashi,
  %``Lattice chiral symmetry and the Wess-Zumino model,''
  Nucl.\ Phys.\  B {\bf 622}, 115 (2002)
  [arXiv:hep-th/0109156].
  %%CITATION = NUPHA,B622,115;%%
  
 %\cite{Fujikawa:2001ns}
\bibitem{Fujikawa:2001ns}
  K.~Fujikawa and M.~Ishibashi,
  %``Lattice chiral symmetry, Yukawa couplings and the Majorana condition,''
  Phys.\ Lett.\  B {\bf 528}, 295 (2002)
  [arXiv:hep-lat/0112050].
  %%CITATION = PHLTA,B528,295;%%

%\cite{Fujikawa:2002ic}
\bibitem{Fujikawa:2002ic}
  K.~Fujikawa,
  %``Supersymmetry on the lattice and the Leibniz rule,''
  Nucl.\ Phys.\  B {\bf 636}, 80 (2002)
  [arXiv:hep-th/0205095].
  %%CITATION = NUPHA,B636,80;%%
  
%\cite{Bonini:2004pm}
\bibitem{Bonini:2004pm}
  M.~Bonini and A.~Feo,
  %``Wess-Zumino model with exact supersymmetry on the lattice,''
  JHEP {\bf 0409}, 011 (2004)
  [arXiv:hep-lat/0402034].
  %%CITATION = JHEPA,0409,011;%%

%\cite{Kikukawa:2004dd}
\bibitem{Kikukawa:2004dd}
  Y.~Kikukawa and H.~Suzuki,
  %``A local formulation of lattice Wess-Zumino model with exact U(1)R
  %symmetry,''
  JHEP {\bf 0502}, 012 (2005)
  [arXiv:hep-lat/0412042].
  %%CITATION = JHEPA,0502,012;%%

%\cite{Bonini:2005qx}
\bibitem{Bonini:2005qx}
  M.~Bonini and A.~Feo,
  %``Exact lattice Ward-Takahashi identity for the N = 1 Wess-Zumino model,''
  Phys.\ Rev.\  D {\bf 71}, 114512 (2005)
  [arXiv:hep-lat/0504010].
  %%CITATION = PHRVA,D71,114512;%%
  
%\cite{Chen:2010uc}
\bibitem{Chen:2010uc}
  C.~Chen, E.~Dzienkowski and J.~Giedt,
  %``Lattice Wess-Zumino model with Ginsparg-Wilson fermions: One-loop results
  %and GPU benchmarks,''
  arXiv:1005.3276 [hep-lat].
  %%CITATION = ARXIV:1005.3276;%%


   %\cite{Osterwalder:1973dx}
\bibitem{Osterwalder:1973dx}
  K.~Osterwalder and R.~Schrader,
  %``AXIOMS FOR EUCLIDEAN GREEN'S FUNCTIONS,''
  Commun.\ Math.\ Phys.\  {\bf 31}, 83 (1973).
  %%CITATION = CMPHA,31,83;%%
  
  %\cite{Osterwalder:1974tc}
\bibitem{Osterwalder:1974tc}
  K.~Osterwalder and R.~Schrader,
  %``Axioms For Euclidean Green's Functions. 2,''
  Commun.\ Math.\ Phys.\  {\bf 42}, 281 (1975).
  %%CITATION = CMPHA,42,281;%%
 
    %\cite{Osterwalder:1977pc}
\bibitem{Osterwalder:1977pc}
  K.~Osterwalder and E.~Seiler,
  %``Gauge Field Theories On The Lattice,''
  Annals Phys.\  {\bf 110}, 440 (1978).
  %%CITATION = APNYA,110,440;%%
  
%\cite{Luscher:1976ms}
\bibitem{Luscher:1976ms}
  M.~Luscher,
  %``Construction Of A Selfadjoint, Strictly Positive Transfer Matrix For
  %Euclidean Lattice Gauge Theories,''
  Commun.\ Math.\ Phys.\  {\bf 54}, 283 (1977).
  %%CITATION = CMPHA,54,283;%%
  
  %\cite{Menotti:1987cq}
\bibitem{Menotti:1987cq}
  P.~Menotti and A.~Pelissetto,
  %``Osterwalder-Schrader Positivity For The Wilson Action,''
  Nucl.\ Phys.\ Proc.\ Suppl.\  {\bf 4}, 644 (1988).
  %%CITATION = NUPHZ,4,644;%%

%\cite{Luscher:1984is}
%\bibitem{Luscher:1984is}
 % M.~Luscher and P.~Weisz,
  %``Definition And General Properties Of The Transfer Matrix In Continuum Limit
  %Improved Lattice Gauge Theories,''
 % Nucl.\ Phys.\  B {\bf 240}, 349 (1984).
  %%CITATION = NUPHA,B240,349;%%

  
  %\cite{Kikukawa:2010gq}
\bibitem{Kikukawa:2010gq}
  Y.~Kikukawa and K.~Usui,
  %``Reflection Positivity of Free Overlap Fermions,''
 Phys.\ Rev.\ D {\bf 82}, 114503 (2010)
 [arXiv:1005.3751 [hep-lat]].
  %%CITATION = ARXIV:1005.3751;%%
 
 %\cite{Luscher:2000hn}
\bibitem{Luscher:2000hn}
  M.~Luscher,
  %``Chiral gauge theories revisited,''
  arXiv:hep-th/0102028.
  %%CITATION = HEP-TH/0102028;%%

 %\cite{Luscher:2000hn}
\bibitem{Usui:2011}
  K.~Usui,
  %``Chiral gauge theories revisited,''
  In preparation.
  %%CITATION = HEP-TH/0102028;%%
 
%\cite{Elliott:2008jp}
\bibitem{Elliott:2008jp}
  J.~W.~Elliott, J.~Giedt and G.~D.~Moore,
  %``Lattice four-dimensional N=4 SYM is practical,''
  Phys.\ Rev.\  D {\bf 78}, 081701 (2008)
  [arXiv:0806.0013 [hep-lat]].
  %%CITATION = PHRVA,D78,081701;%%

  
%\cite{Kikukawa:2002as}
\bibitem{Kikukawa:2002as}
  Y.~Kikukawa and Y.~Nakayama,
  %``Nicolai mapping vs. exact chiral symmetry on the lattice,''
  Phys.\ Rev.\  D {\bf 66}, 094508 (2002).
%  [arXiv:hep-lat/0207013].
  %%CITATION = PHRVA,D66,094508;%%

%\cite{Kawai:2010yj}
\bibitem{Kawai:2010yj}
  H.~Kawai and Y.~Kikukawa,
  %``A Lattice study of N=2 Landau-Ginzburg model using a Nicolai map,''
  Phys.\ Rev.\  D {\bf 83}, 074502 (2011)
  [arXiv:1005.4671 [hep-lat]].
  %%CITATION = PHRVA,D83,074502;%%
  
  %\cite{Kikukawa:2008xw}
\bibitem{Kikukawa:2008xw}
  Y.~Kikukawa and F.~Sugino,
  %``Ginsparg-Wilson Formulation of 2D N =(2,2) SQCD with Exact Lattice
  %Supersymmetry,''
  Nucl.\ Phys.\  B {\bf 819}, 76 (2009)
  [arXiv:0811.0916 [hep-lat]].
  %%CITATION = NUPHA,B819,76;%%

  %\cite{Kato:2008sp}
\bibitem{Kato:2008sp}
  M.~Kato, M.~Sakamoto and H.~So,
  %``Taming the Leibniz Rule on the Lattice,''
  JHEP {\bf 0805}, 057 (2008). 
%  [arXiv:0803.3121 [hep-lat]].
  %%CITATION = JHEPA,0805,057;%%

% fixed point approach

%\cite{Hasenfratz:1998ri}
\bibitem{Hasenfratz:1998ri}
P.~Hasenfratz, V.~Laliena and F.~Niedermayer,
%``The index theorem in QCD with a finite cut-off,''
Phys.\ Lett.\ B {\bf 427}, 125 (1998)
[arXiv:hep-lat/9801021].
%%CITATION = HEP-LAT 9801021;%%

%\cite{Hasenfratz:1998jp}
\bibitem{Hasenfratz:1998jp}
P.~Hasenfratz,
%``Lattice QCD without tuning, mixing and current renormalization,''
Nucl.\ Phys.\ B {\bf 525}, 401 (1998)
[arXiv:hep-lat/9802007].
%%CITATION = HEP-LAT 9802007;%%


%\cite{Bell-Wilson:1975}
\bibitem{Bell-Wilson:1975}
T.~L.~Bell and K.~G.~Wilson,  
%``Finite-lattice approximations to renormalization groups,''
Phys. Rev. {\bf B11} (1975)  3431 

%\cite{Wiese:1993cb}
\bibitem{Wiese:1993cb}
  U.~J.~Wiese,
  %``Fixed point actions for Wilson fermions,''
  Phys.\ Lett.\  B {\bf 315}, 417 (1993)
  [arXiv:hep-lat/9306003].
  %%CITATION = PHLTA,B315,417;%%

%\cite{So:1998ya}
\bibitem{So:1998ya}
  H.~So and N.~Ukita,
  %``Ginsparg-Wilson relation and lattice supersymmetry,''
  Phys.\ Lett.\  B {\bf 457}, 314 (1999)
  [arXiv:hep-lat/9812002].
  %%CITATION = PHLTA,B457,314;%%
\end{thebibliography}
\end{document}